\documentclass[
preprint,
prb,
superscriptaddress,
showpacs,
nofootinbib,
amsmath,
amssymb,
aps,
longbibliography,
]{revtex4-2}
\usepackage{framed}
\usepackage{graphicx}
\usepackage{dcolumn}
\usepackage{amsmath}
\usepackage{amssymb}
\usepackage{color}
\usepackage{subfigure,amsmath,verbatim,moreverb,bm}

\def\bef{\begin{framed}}
\def\eef{\end{framed}}
\def\be{\begin{equation}}
\def\ee{\end{equation}}
\def\ber{\begin{eqnarray}}
\def\eer{\end{eqnarray}}

\def\nablav{{\boldmath{\nabla}}}

\def\Omegav{\mbox{\boldmath $\Omega$}}

\def\nuv{\mbox{\boldmath $\nu$}}

\def\rv{{\bf r}}

\def\fv{{\bf f}}

\def\Mv{{\bf M}}
\def\mv{{\bf m}}

\def\zv{{\bf \hat z}}

\def\vv{{\bf v}}

\def\xv{{\bf \hat x}}
\def\jv{{\bf j}}

\def\kv{{\bf k}}
\def\qv{{\bf q}}
\def\Av{{\bf A}}
\def\Bv{{\bf B}}

\def\vv{{\bf v}}
\def\nn{\nonumber}

\begin{document}
\title{Ward identities and orbital magnetization in current density functional theory}
\author{Giovanni Vignale}
\email{vgnl.g@nus.edu.sg}
\affiliation{The Institute for Functional Intelligent Materials (I-FIM), National University of Singapore, 4 Science Drive 2, Singapore 117544}
\author{Junren Shi}
\affiliation{International Center for Quantum Materials, Peking University, Beijing 100871, China}
\author{Di Xiao}
\affiliation{Department of Materials Science and Engineering, University of Washington, Seattle, Washington 98195, USA}
\affiliation{Department of Physics, University of Washington, Seattle, WA 98195, USA}

\author{Qian Niu}
\affiliation{School of Physics, University of Science and Technology of China, Hefei, Anhui 230026, China}

\date{\today}

\begin{abstract}
We revisit the derivation of the orbital magnetization formula for periodic crystals in  current density functional theory (CDFT)~\cite{Shi2007}. Our new derivation computes the linear response of the energy density to a periodic magnetic field in the long-wavelength limit.  We unveil a   Ward identity which connects the current vertex to the derivative of the Kohn-Sham self-energy. The result of Ref.~\cite{Shi2007} is confirmed: the orbital magnetization of the interacting solid can be computed exactly (in principle) from the self-consistent eigenfunctions and eigenvalues of the Kohn-Sham equation of CDFT. 
\end{abstract}
\maketitle

\newpage
\section{Introduction}\label{Introduction}
The ``modern theory" of electronic orbital magnetization for a periodic solid is by now well established~\cite{Chang1996,Thonhauser2005,DiXiao2005,Ceresoli2006,DiXiao2006,Shi2007,Chang_2008,NiuXiao2010,Vanderbilt_2018,Aryasetiawan2019}. For a system described by Bloch wave functions $\psi_{n,\kv}(\rv)=e^{i\kv\cdot\rv}u_{n,\kv}(\rv)$ and band energies $\epsilon_{n,\kv}$, where $n$ is a band index and $\kv$ a Bloch wave vector in the first Brillouin zone, the magnetization is given by

\be\label{ModernTheory}
\Mv= \sum_{n,\kv} f_{n,\kv}\mv_{n,\kv}+\frac{e}{\beta \hbar}\sum_{n,\kv}\Omegav_{n,\kv} \ln{\left[1+e^{-\beta(\epsilon_{n,\kv}-\mu)}\right]}\,.
\ee
where $\mu$ is the chemical potential, $\beta=1/(k_BT)$ the inverse temperature, $f_{n,\kv}$ the Fermi-Dirac distribution at energy $\epsilon_{n,\kv}$ and the Berry curvature and the orbital moment are defined in terms of the periodic parts of the Bloch wave functions as follows:
\be\label{BerryCurvature}
\Omegav_{n,\kv}=i\langle\nablav_\kv u_{n,\kv}|\times|\nablav_\kv u_{n,\kv}\rangle\,.
\ee
and
\be\label{OrbitalMoment}
\mv_{n,\kv}
=\frac{ie}{2\hbar }\langle \nablav_\kv u_{n,\kv}|[\epsilon_{n,\kv}-\hat H_{\kv}] \times|\nablav_\kv u_{n,\kv}\rangle
\,,
\ee
$\hat H_{\kv}$ being the one-electron Hamiltonian in the $\kv$-sector of the Hilbert space.
At the absolute zero of temperature, Eq.~(\ref{ModernTheory}) simplifies to
\be\label{ModernTheory0}
\Mv^{(0)}= \frac{i e}{2\hbar}\sum_{n,\kv}f^{(0)}_{n,\kv}\langle \nablav_\kv u_{n,\kv}|[\epsilon_{n,\kv}+\hat H_\kv-2\mu] \times|\nablav_\kv u_{n,\kv}\rangle
\,,
\ee
where $f^{(0)}_{n,\kv}$ is the zero-temperature Fermi-Dirac distribution.

The above equations were originally derived for noninteracting electrons, and it is not immediately obvious that they remain valid in interacting systems. One way to introduce interactions is to replace the non-interacting Hamiltonian by an effective mean-field Hamiltonian, for example the Hartree-Fock (HF) Hamiltonian, and similarly replace the Bloch eigenvalues and eigenfunctions by self-consistent HF eigenvalues and eigenfunctions.  Indeed, this has been recently shown to be the correct procedure  within HF theory~\cite{Vafek2025,Chunli2025,Xiao2025}. However, the resulting magnetization will suffer from  the well known limitations of the HF approximation, stemming from the neglect of correlation.

Density functional theory (DFT)~\cite{HK64,KS65,Grayce94} offers a more appealing solution since it is a formally exact theory, yet easier to use than the HF approximation. The problem with ordinary DFT is that its ``formal exactness" is limited in scope: in particular, the Kohn-Sham equation of ordinary DFT does not give access to the exact orbital current density, nor to the exact orbital magnetization.

Fortunately, a generalization of ordinary DFT exists -- the current-density functional theory (CDFT)\cite{VR87,VR88,GV2005} --  which gives direct access to the exact orbital current density via Kohn-Sham single-particle wave functions. This strongly suggests that Eqs.~(\ref{ModernTheory}) and (\ref{ModernTheory0}) can be made {\it formally exact} through the use of eigenvalues and eigenfunctions obtained from the self-consistent solution of the exact Kohn-Sham equation of CDFT.

Indeed, this is the conclusion that we  reached in Ref.\cite{Shi2007} even though, in hindsight, we see that the analysis was not carried out in sufficient depth. Going back to the arguments presented in Ref.\cite{Shi2007} we have identified two issues: 

(1) The CDFT part of Ref.\cite{Shi2007} relied on calculating the derivative of the {\it total energy} with respect to a uniform magnetic field: this is misaligned with the first part of the paper, in which we computed the derivative of the {\it energy density} with respect to a periodic magnetic field (in the long-wavelength limit). While the total energy approach is in principle valid,  it is now  understood that it faces significant difficulties in the treatment of surface/edge state contributions, especially so in the case of nontrivial Chern insulators~\cite{Thonhauser2005,Ceresoli2006}.  {By contrast, the energy density approach works entirely with bulk wave functions and correctly includes the spectral flow of the density of states in a magnetic field.}

(2) After conflating the energy density approach with the total energy approach, we went on to argue that Eq.~(\ref{ModernTheory}) would remain unchanged in the CDFT/Kohn-Sham framework because the variation of the self-consistent exchange-correlation potentials (induced by the externally applied magnetic field) does not contribute, by virtue of the variational principle, to the first-order change of the total energy. This appears to make the Kohn-Sham system formally equivalent to a genuine non-interacting system, i.e., a system in which only the external vector potential changes as a result of the magnetic perturbation. This argument overlooks the fact that the external potential couples to the {\em physical velocity}, defined as the commutator of the position operator with the noninteracting Hamiltonian (see discussion after Eq.~(\ref{Vprime})), but this differs from the ``Kohn-Sham velocity" -- defined as the commutator of the position operator with the Kohn-Sham Hamiltonian -- by an ``exchange-correlation" (xc)  vector potential term~\cite{VR87}. This breaks the formal equivalence between the Kohn-Sham system and the noninteracting system.  

Having recognized these issues (and motivated in part by Refs.~\cite{Vafek2025,Chunli2025,Xiao2025}), we have carefully re-examined the claim of Ref.[1].  Happily, we can confirm that our conclusion was right: the overlooked features compensate each other and the final upshot is that the exact Kohn-Sham eigenfunctions and eigenvalues of CDFT do provide the correct orbital magnetization of the interacting system, when plugged in Eq.~(\ref{ModernTheory}).

A crucial role in the proof is played by a new set of Ward identities~\cite{Nozieres64,Schrieffer}, which connect the renormalization of the current
vertex (arising from the difference between the Kohn-Sham velocity and the physical velocity) to the eigenvalues of the Kohn-Sham equation. The proof of these identities will be presented in section \ref{WardIdentity}.

\section{Orbital magnetization in a noninteracting system}\label{NonInteractingSystem}

We begin by reviewing the calculation of the orbital magnetization in a noninteracting system. We work, for simplicity, at zero temperature. We consider a perfectly periodic crystal with a very large number of unit cells and periodic boundary conditions a la Born-von Karman (BvK). Let us assume that the crystal has an average ground state magnetization $\Mv$.  This can only happen if time-reversal symmetry is broken, either spontaneously, or by the application of a magnetic field $\Bv$.  In the former case one must still assume that an infinitesimal uniform $\Bv$ is present to pin the orientation of $
\Mv$.   We apply a small periodic magnetic field of amplitude $\delta\Bv_\qv$ causing a perturbation 
\be\label{MagneticPerturbation}
\hat V_B = \hat \jv_{-\qv}\cdot \delta\Av_\qv
\ee
where $\hat \jv_{-\qv}$ is the current density fluctuation operator at wave vector $-\qv$ (see Eq.~(\ref{CurrentDensityFluctuation}) below) and  $\delta\Av_\qv$ is the amplitude of the periodic vector potential that generates $\delta \Bv_\qv$.\footnote{For a magnetic field along the $\zv$ axis ($\delta \Bv_\qv=\delta B_\qv\zv$) and $\qv$ along the $\xv$-axis ($\qv = q\xv$) we choose
\be\label{deltaAq}
\delta\Av_\qv = \frac{\delta B_\qv}{2iq}{\bf\hat y}\,.
\ee
Here $\qv$ is a small wave vector (much smaller than the Brillouin zone) which tends to zero while satisfying BvK periodic boundary conditions.}
The periodic magnetic field combines with the uniform magnetization to produce a periodic (grand-canonical) energy density at the same wave vector
\be
\delta K_\qv = -\Mv\cdot \delta \Bv_\qv\,.
\ee
Thus, $\Mv$ is (minus) the coefficient of proportionality between $\delta K_\qv$ and $\delta\Bv_\qv$. 

The calculation of the grand-canonical energy density fluctuation at wave vector $\qv$ for a noninteracting system has been described in Ref.[1], but here we use a diagrammatic approach which is easier to generalize to the interacting case.
We introduce the noninteracting energy density fluctuation operator
\be
\hat K_{0,\qv} =\frac{1}{2}\sum_i \left\{\frac{1}{2}[-i\nabla_i +\Av(\hat\rv_i)]^2+V(\hat\rv_i)-\mu,e^{-i\qv\cdot\hat \rv_i}\right\}
\ee
where the sum runs over the particles and $\left\{..,..\right\}$ is the anticommutator ($V$ is the periodic crystal potential, $\Av$ is the vector potential of a uniform magnetic field pre-existing the perturbation), $\mu$ is the chemical potential.
We work in atomic units $e=\hbar=m=1$.
The current density fluctuation operator is
\be\label{CurrentDensityFluctuation}
\hat \jv_{-\qv} =\frac{1}{2} \sum_i \left\{\hat\vv_i,e^{i\qv\cdot\hat \rv_i}\right\}\,,
\ee
where $\hat \vv_i$ is the physical velocity operator,  defined as the commutator of the position operator of the $i$-th particle with the Hamiltonian. 
\begin{figure}[ht]
\centering
\includegraphics[scale=0.5]{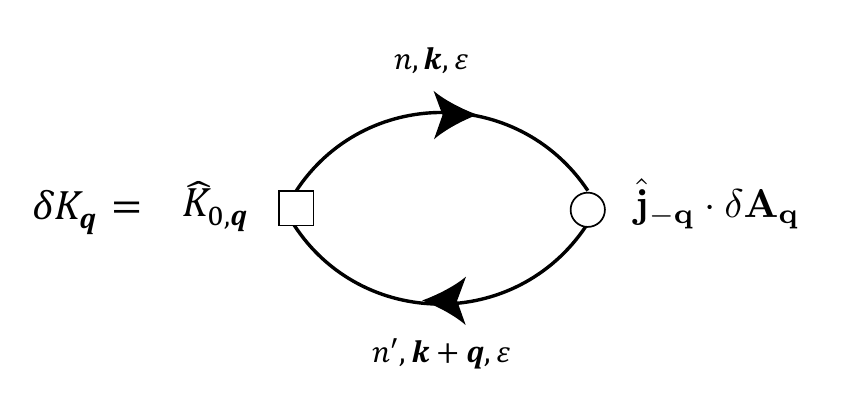}
\caption{\footnotesize{Bubble diagram for the response of the noninteracting energy density fluctuation to a periodic magnetic field. The current density fluctuation operator is defined in Eq.~(\ref{CurrentDensityFluctuation}).}}
\label{Fig1}
\end{figure} 
The response of the energy density fluctuation to the periodic magnetic perturbation is given by the ``bubble diagram" of Fig.~\ref{Fig1}. The solid lines are noninteracting Green's functions associated with quantum numbers $n',\kv+\qv$ and $n,\kv$, both evaluated at the  frequency $\varepsilon$:
\be\label{GreenFunction0}
G_{n,\kv}(\varepsilon)=\frac{1}{\varepsilon -\epsilon_{n,\kv}+i\eta ~{\rm sgn}(\varepsilon-\mu)}\,.
\ee
Notice that the Green's function is diagonal in both band and momentum indices. The square vertex represents the matrix element of $\hat K_{0,\qv}$ between the outgoing and the incoming states. This matrix element is explicitly given by
\be\label{H0Q}
[\hat K_{0,\qv}]_{n\kv,n'\kv+\qv}=\frac{1}{2}\left(\epsilon_{n,\kv}+\epsilon_{n',\kv+\qv}-2\mu\right)\langle u_{n,\kv}|u_{n',\kv+\qv}\rangle\,.
\ee
The circle vertex represents the matrix element of $\hat \jv_{-\qv}$ between the outgoing and the incoming states.  Its explicit form is\footnote{This expression is valid in the ``parallel transport" gauge, in which $\langle u_{n,\kv}|\nablav_\kv u_{n,\kv}\rangle=0$ at the given value of $\kv$.}
\be\label{J-Q}
[\hat j_{-\qv}]_{n'\kv+\qv,n\kv}=\left(\nablav_\kv\epsilon_{n,\kv}\right)\langle u_{n,\kv+\qv}|u_{n,\kv}\rangle \delta_{n,n'}+i\left(\epsilon_{n',\kv+\qv}-\epsilon_{n,\kv}\right)\langle u_{n',\kv+\qv}|\nablav_\kv u_{n\kv}\rangle (1-\delta_{n,n'})\,.
\ee

{Evaluation of the diagram by the standard rules\cite{GV2005} gives 
\be
M_z  = -\lim_{q\to0}\frac{1}{q}\sum_{n,n',\kv}\frac{f^{(0)}_{n',\kv+q\xv}-f^{(0)}_{n,\kv}}{\epsilon_{n',\kv+q\xv}-\epsilon_{n,\kv}}\Im m 
\left\{[\hat K_{0,\qv}]_{n\kv,n'\kv+\qv}~[\hat j_{-\qv}]_{n'\kv+\qv,n\kv}\right\}\,,
\ee
which, after algebraic manipulations,  reduces to the zero-temperature  noninteracting magnetization formula~(\ref{ModernTheory0}). 
It should be noted that the matrix element of the energy density fluctuation operator $[\hat K_{0,\qv}]_{n\kv,n'\kv+\qv}$ vanishes in the limit $q\to 0$, canceling the divergence of $1/q$.  For this reason, the linear in $\qv$ part of the current vertex does not contribute to the magnetization.}

\section{Orbital magnetization in CDFT}\label{CDFT-OrbitalMagnetization}
To calculate the energy density response in CDFT (working at $T=0$ for simplicity) we start from the ground state energy functional~\cite{VR87,GV2005}
\be
K[n,\jv_p] =T_{s,\Av}[n,\jv_p]+\int n(\rv) [V(\rv)-\mu] d\rv +E_{Hxc}[n,\jv_p]\,,
\ee 
where $n$ and $\jv_p$ are the density and the paramagnetic current density in the ground state\footnote{The paramagnetic current density $\jv_p$  differs from the physical current density $\jv$ by not including the vector potential contribution to the velocity, i.e., $\jv=\jv_p+n\Av$.}.
The noninteracting gauge-invariant kinetic energy functional, $T_{s,\Av}[n,\jv_p]$,  is expressed in terms of KS orbitals
\be
T_{s,\Av}[n,\jv_p]=\sum_{n,\kv}f^{(0)}_{n,\kv}\langle\psi_{n,\kv}|\frac{1}{2}[-i\nablav_\rv +\Av(\hat\rv)]^2|\psi_{n,\kv}\rangle\,,
\ee
where $|\psi_{n,\kv}\rangle$ are single-particle orbitals uniquely determined by the densities $n$ and $\jv_p$ according to the Hohenberg-Kohn theorem of CDFT~\cite{VR87}.
$E_{Hxc}[n,\jv_p]$ is the Hartree + exchange + correlation energy functional. Again, $\Av(\rv)$ is the vector potential of a uniform magnetic field, which pre-exists the application of the magnetic perturbation.  Similarly, the density and the paramagnetic current density are expressed in terms of KS orbitals as
\be\label{Density}
n(\rv)=\sum_{n,\kv}f^{(0)}_{n,\kv}|\psi_{n,\kv}(\rv)|^2
\ee
and 
\be\label{PCurrentDensity}
\jv_p(\rv)=\sum_{n,\kv}f^{(0)}_{n,\kv}\Im m  \left[\psi^*_{n,\kv}(\rv)\nablav_\rv \psi_{n,\kv}(\rv)\right]\,.
\ee
The ground state energy functional can now be written as
\be\label{EnergyFunctional}
K[n,\jv_p] = \int d\rv \sum_{n,\kv}f^{(0)}_{n,\kv}\psi^*_{n,\kv}(\rv) \hat K_0\psi_{n,\kv}(\rv) +E_{Hxc}[n,\jv_p]
\ee
where
\be\label{NonInteractingHamiltonian}
 \hat K_0=\frac{1}{2}[-i\nabla +\Av(\hat\rv)]^2+V(\hat\rv)-\mu
 \ee
 is the noninteracting  grand-canonical Hamiltonian.

 Requiring the energy functional to be stationary for small variations of $n$ and $\jv_p$ at constant external potentials gives the KS equation of CDFT\cite{VR87,VR88}:
\be\label{KohnShamEquation}
\left[\frac{1}{2}\left(-i\nabla +\Av'\right)^2+V'\right]\psi_{n,\kv}(\rv)=\epsilon_{n,\kv} \psi_{n,\kv}(\rv)\,,
\ee
\be\label{Aprime}
\Av'(\rv)=\Av(\rv)+\Av_{xc}(\rv)\,,
\ee
\be\label{Vprime}
V'(\rv)=V(\rv)+V_{Hxc}(\rv)-\Av(\rv)\cdot\Av_{xc}(\rv)-\frac{1}{2}A_{xc}^2(\rv)\,,
\ee
where $\Av_{xc}(\rv) \equiv \left.\frac{\delta E_{Hxc}[n,\jv_p]}{\delta \jv_p(\rv)}\right\vert_{n}$ 
is the xc vector potential and $V_{Hxc}(\rv)\equiv \left.\frac{\delta E_{Hxc}[n,\jv_p]}{\delta n(\rv)}\right\vert_{\jv_p}$ is the sum of Hartree and xc potentials.\footnote{Eq.~(\ref{KohnShamEquation}) is not {\it manifestly} gauge-invariant, because $E_{Hxc}[n,\jv_p]$ is expressed in terms of the non-gauge-invariant variable $\jv_p$.  A manifestly gauge-invariant Hxc functional can be constructed in terms of the vorticity $\nuv(\rv) \equiv \nablav_\rv\times \frac{\jv_p(\rv)}{n(\rv)}$~\cite{VR87} -- in other words we set $E_{Hxc}[n,\jv_p]=\bar E_{xc}[n,\nuv]$.
With this definition, we get
 \be 
\left(\frac{1}{2}\left[(-i\nablav+\Av)^2 + \left\{-i\nablav-\frac{\jv_p}{n}, \Av_{xc}\right\}\right]+ V+\bar V_{Hxc}\right)\psi_{n,\kv}(\rv)=\epsilon_{n,\kv} \psi_{n,\kv}(\rv)\nonumber
\ee
where the curly bracket $\{A,B\}=AB+BA$ denotes the anticommutator. 
The {\it gauge-invariant potential} $\bar V_{Hxc}$ is defined as  $\bar V_{Hxc}=\delta\bar E_{Hxc}[n,{\nuv}]/\delta n|_{\nuv}$.  The relation between  $\bar V_{Hxc}$ and $V_{Hxc}$ is 
\be
V_{Hxc}=\bar V_{Hxc} - \Av_{xc}\cdot\frac{\jv_p}{n}\,.
\ee
}

 Just as in the noninteracting case, the magnetic perturbation is introduced by adding to the external vector potential a small periodic component of amplitude $\delta\Av_\qv$. This modifies both $\Av'$, Eq.~(\ref{Aprime}), and $V'$, Eq.~(\ref{Vprime}). Remarkably, the two terms conspire to produce a magnetic perturbation that is still given by Eq.~(\ref{MagneticPerturbation}), where $\hat\jv_{-\qv}$ is the {\em bare} physical current. The presence of $\Av_{xc}$ does not modify the form of the magnetic perturbation!
 
 Besides explicitly modifying the Kohn-Sham Hamiltonian, the magnetic perturbation induces changes $\delta n(\rv)$ and $\delta\jv_p(\rv)$ in the densities.  The corresponding change in the energy, following from Eq.~(\ref{EnergyFunctional}) is
 \ber\label{deltaE}
 \delta K&=&\delta\int d\rv \sum_{n,\kv}f^{(0)}_{n,\kv}\psi^*_{n,\kv}(\rv) \hat K_0\psi_{n,\kv}(\rv)\nn\\
 &+& \int d\rv V_{Hxc}(\rv)\delta n(\rv)+\int d\rv \Av_{xc}(\rv)\cdot \delta\jv_p(\rv)\,,
 \eer 
 where $V_{Hxc}(\rv)$ and $\Av_{xc}(\rv)$ are the effective potentials evaluated at the unperturbed densities (i.e., the densities before applying the magnetic perturbation).  The $\delta$ on the first line includes the variations of $\hat K_0$, $\psi_{n,\kv}$ and $f^{(0)}_{n,\kv}$.   The first and second lines of Eq.~(\ref{deltaE}) can be combined in a single variation
 \be\label{deltaE2}
\delta K =\int d\rv~  \bar \delta\left\{\sum_{n,\kv}f^{(0)}_{n,\kv}\psi^*_{n,\kv}(\rv) \hat K_{KS}\psi_{n,\kv}(\rv)\right\}\,,
\ee
  where $\hat H_{KS}$ is the KS hamiltonian 
  \be
  \hat K_{KS}\equiv \hat K_0+\hat V_{Hxc}+\hat A_{xc}\,,
  \ee
  where we have defined the operators
  \be
   \hat V_{Hxc}\equiv V_{Hxc}(\hat \rv)
   \ee
   and
    \be
   \hat A_{xc}\equiv \frac{1}{2}\left\{ -i\nablav \cdot \Av_{xc}(\hat\rv)+\Av_{xc}(\hat\rv)\cdot (-i\nablav)\right\}\,,
    \ee
    with effective potentials $V_{Hxc}$ and $\Av_{xc}$ evaluated at the unperturbed densities.
   {It is essential for Eq.~(\ref{deltaE2}) to hold true that the Hxc potentials within the operator $\hat K_{KS}$ remain fixed under the variation, pinned at their equilibrium values.  At the same time, the variation of the wave functions $\psi_{n,\kv}$ and their eigenvalues   must fully include the effect of the varying  $Hxc$ following from varying densities.  The notation $\bar \delta$, as opposed to $\delta$,  in Eq.~(\ref{deltaE2}) is introduced as a reminder of this important fact. Notice that the  first order variation of $\hat K_0$, due to the magnetic perturbation,  produces no change in the energy, since a $\qv-$periodic perturbation has zero expectation value in a Bloch wave function. Therefore the meaning of $\bar \delta$ is best explained by saying that  $\hat K_{KS}$ remains constant in the variation of  Eq.~(\ref{deltaE2}).}

The quantity under the integral of Eq.~(\ref{deltaE2}) is naturally interpreted as the first-order variation of the {\it energy density}. 
It may be objected that the energy density (unlike the total energy) cannot be legitimately computed from the Kohn-Sham equation. This concern is addressed as follows. The CDFT energy density at point $\rv$ can be constructed by considering the {\it total energy} response to local perturbations (i.e., scalar and vector potentials) which modify the values of density and current only in a small region of space around a point $\rv$. By the principle of nearsigthedness~\cite{Prodan2005}, the resulting change in total energy  can be interpreted as the change of the energy density at point $\rv$. To the extent that the change in total energy can be regarded as ``formally exact", this establishes the formal exactness of the energy density at point $\rv$. 

\begin{figure}[ht]
\centering
\includegraphics[scale=0.5]{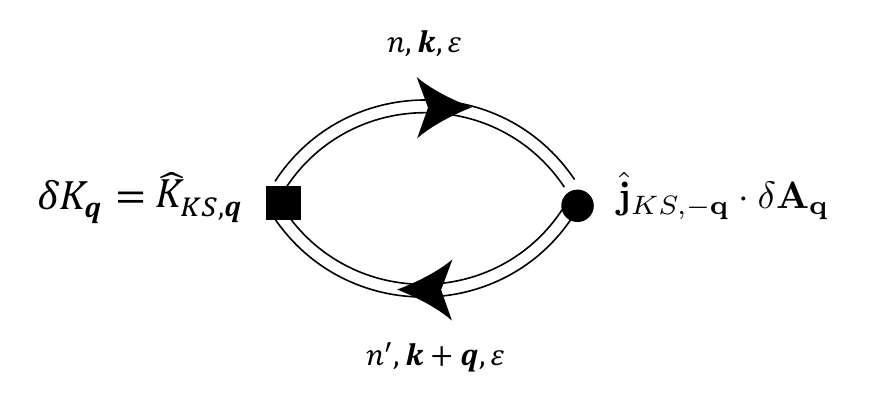}
\caption{\footnotesize{Bubble diagram for the response of the KS energy density fluctuation (from CDFT) to a periodic magnetic field.}}
\label{Fig2}
\end{figure} 

Fig.~\ref{Fig2} shows the bubble diagram for the response of the KS energy density fluctuation to the magnetic perturbation.
Comparing this to Fig.~\ref{Fig1}, we see that three things have changed.
\begin{enumerate}
\item
The vertex for the noninteracting energy density fluctuation is replaced by the corresponding quantity for the KS energy density fluctuation (black square).
This is still given by Eq.~(\ref{H0Q}) except that the band energies and the periodic eigenstates are now obtained self-consistently from the KS equation.
\item 
The noninteracting Green's function is replaced by a KS Green's function (double line):
\be\label{GreenFunctionKS}
G^{KS}_{n,\kv}(\varepsilon)=\frac{1}{\varepsilon -\epsilon^{KS}_{n,\kv}+i\eta ~{\rm sgn}(\varepsilon-\mu)}\,,
\ee
where $\epsilon^{KS}_{n,\kv}$ are KS energies.

\item The bare current vertex is replaced by the ``dressed" current vertex (black circle), which is constructed from the matrix elements of the ``KS velocity",
\be\label{KSVelocity}
\hat \vv_{KS,\kv} \equiv i[\hat H_{KS,\kv},\hat \rv]= \frac{\partial \hat H_{KS,\kv}}{\partial \kv}.
\ee
Eq.~(\ref{J-Q}) is still valid, with band energies and periodic states obtained from the KS equation.
\end{enumerate}

The first and second change are fairly obvious. The only delicate point is the replacement of the bare current vertex by the dressed current vertex, which we now discuss in detail. 

\begin{figure}[ht]
\centering
\includegraphics[scale=0.5]{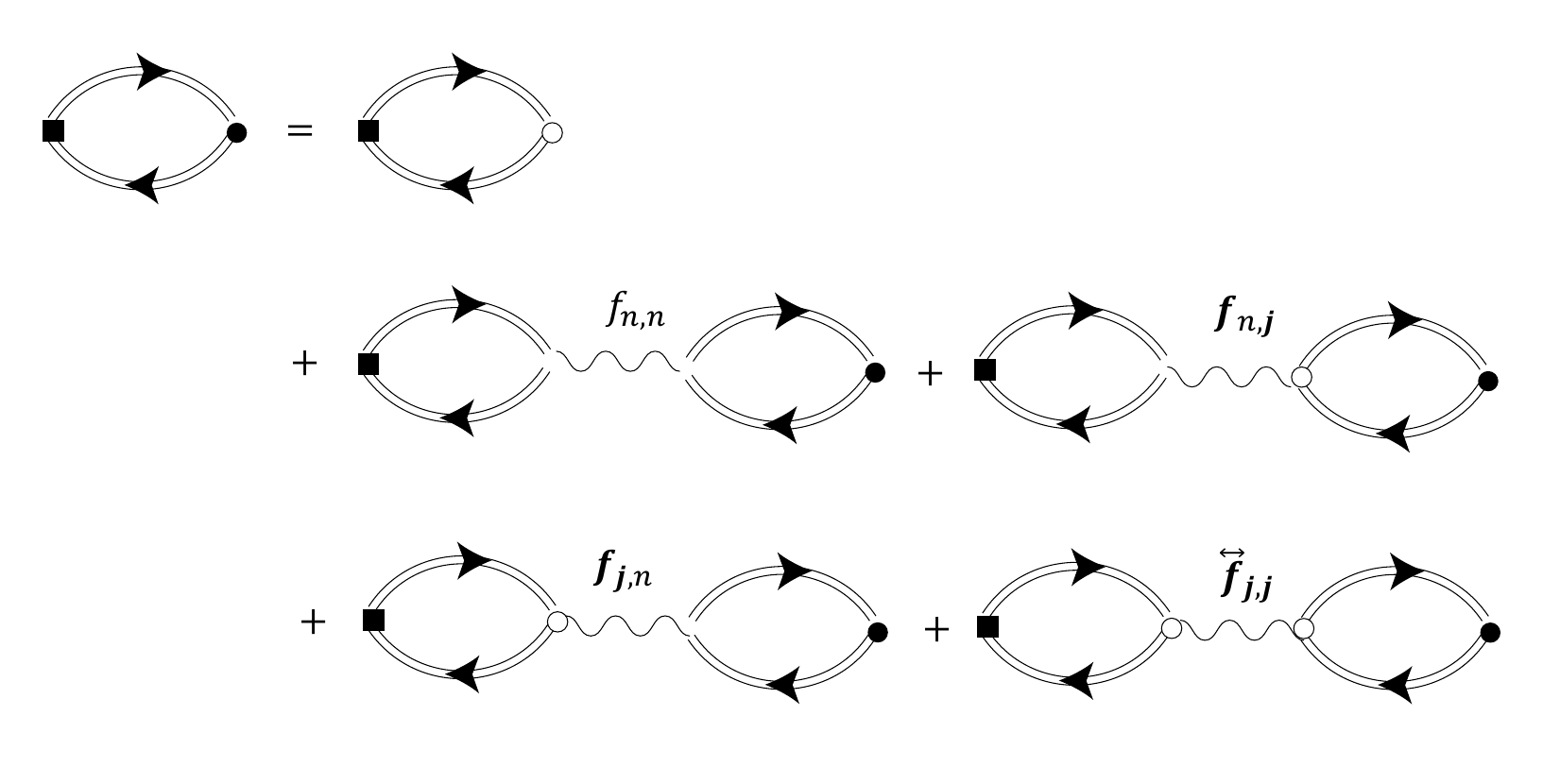}
\caption{\footnotesize{The complete set of diagrams for the response of the energy density.  The bare vertex (white circle) would take care only of the first diagram. In reality, there are four additional families of diagrams controlled by the exchange-correlation kernels $f_{n,n}$, $\fv_{n,\jv}$, $\fv_{\jv,n}$,   and $\stackrel{\leftrightarrow}{\fv}_{\jv,\jv}$, which are represented by wavy lines connecting two pairs of propagators. These kernels describe the response of the effective potentials as shown in Eqs.~(\ref{DeltaVxc}) and (\ref{DeltaAxc}). Notice that the kernels carry the infinitesimal momentum $\qv$, so in the $\qv \to 0$ limit they reduce to numbers such as $\int f_{n,n}(\rv) d\rv$ etc...}}
\label{Fig3}
\end{figure} 

The need to replace the bare vertex by the ``dressed" vertex arises because the bare vertex alone fails to include an infinite series of  RPA-like diagrams which contribute to the magnetic response.  These diagrams are shown in Fig.\ref{Fig3}. They arise because the magnetic perturbation induces density and current fluctuations, which in turn induce fluctuations in the effective potentials $V_{xc}$ and $\Av_{xc}$: these fluctuations contribute to the total response.

Notice that the diagrams in Fig. 3 involve the ``xc kernels" $f_{n,n}$, $\fv_{n,\jv}$, $\fv_{\jv,n}$,   and $\stackrel{\leftrightarrow}{\fv}_{\jv,\jv}$.  These kernels describe the infinitesimal variations of the effective potentials arising from an infinitesimal variation of the densities, and are defined by the relations
\be\label{DeltaVxc}
\delta V_{Hxc}(\rv)=\int f_{n,n}(\rv-\rv')\delta n(\rv') d\rv'  + \int \fv_{n,\jv}(\rv-\rv')\cdot \delta \jv_p(\rv') d\rv'
\ee
and
\be\label{DeltaAxc}
\delta \Av_{xc}(\rv)=\int \fv_{\jv,n}(\rv-\rv')\delta n(\rv') d\rv'  + \int \stackrel{\leftrightarrow}{\fv}_{\jv,\jv}(\rv-\rv')\cdot \delta \jv_p(\rv') d\rv'\,.
\ee
The exchange-correlation kernels, represented by wavy lines in Fig.~\ref{Fig3}, inherit the translational invariance of the Coulomb interaction and are therefore functions of $\rv-\rv'$. 
Each kernel can be represented in the form $f(\rv-\rv')=\sum_\qv f(\qv) e^{i\qv\cdot(\rv-\rv')}$  where $\qv$ is the momentum flowing in the wavy line, and the sum over $\qv$ runs over all wave vectors. 

These kernels give rise to an effective two-body interaction
\be
\hat F = \sum_{\qv}\left\{\hat n_{-\qv}f_{n,n}(\qv) \hat n_\qv+ \hat \jv_{-\qv}\cdot\fv_{\jv,n} (\qv)\hat n_\qv + \hat n_{-\qv} \fv_{n,\jv}(\qv)\cdot\jv_{\qv}+\hat \jv_{-\qv}\cdot  \stackrel{\leftrightarrow}{\fv}_{\jv,\jv} (\qv)\cdot \hat \jv_{\qv}
\right\}\,
\ee
where the density fluctuation vertex ($n_\qv$) is denoted in Fig.~\ref{Fig3} by a blank, and the bare current vertex ($\jv_{\qv}$)  by an open circle. 

The interaction conserves the Bloch wave vector, i.e., if the momentum of the propagator changes from $\kv$ to $\kv+\qv$ at the left end of the wavy line, then it must change by the opposite amount,  $\kv'+\qv$ to $\kv'$, at the right end. Crucially, $f(\qv)$ does not depend on the Bloch wave vectors of the incoming and outgoing particles, but only on the momentum transfer.

 It is not immediately obvious that the exchange-correlation diagrams of Fig. 3, in the limit $\qv\to0$,  are completely accounted for by the replacement $\hat\jv_{-\qv} \to \hat\jv_{KS,-\qv}$, as claimed in Fig.~\ref{Fig2}.  This important result is proved in the next section.
 
 \section{The Ward identity in CDFT}\label{WardIdentity}
\begin{figure}[ht]
\centering
\includegraphics[scale=0.5]{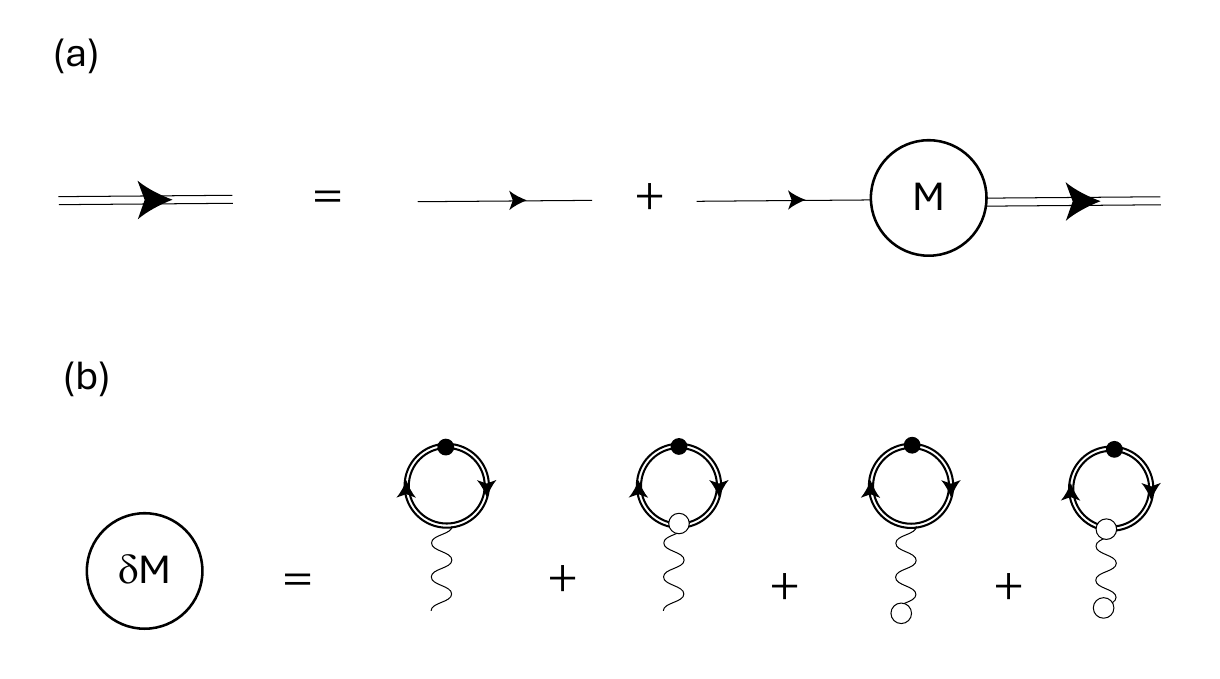}
\caption{\footnotesize{(a) Diagrammatic representation of the Dyson equation (\ref{DysonEquation}) for the KS propagator. (b) The four contributions to the linearized self-energy insertion $\delta M$. Exchange-correlation kernels $f_{n,n}$, $\fv_{n,\jv}$, $\fv_{\jv,n}$ and $\stackrel{\leftrightarrow}{\fv}_{\jv,\jv}$ from left to right are represented by wavy lines. The black dot on the double line denotes the variation, $\delta G_{KS}$, of the KS propagator.}}
\label{Fig4}
\end{figure}

 It is well known that the KS propagator can be expanded in a Dyson series of noninteracting propagators and potential insertions:~\cite{Nozieres64,GV2005}
 \be\label{DysonEquation}
 G_{KS}=G_0+G_0 M G_0+G_0 M G_0M G_0 +... =G_0+G_0 M G_{KS}
 \ee
 where the Kohn-Sham ``self-energy" $M$ is the sum of the $V_{Hxc}$ and $\Av_{xc}$ local potential insertions (see Fig.~\ref{Fig4}). Notice that $G$, $G_0$, and $M$ are matrices with Bloch indices, and  products like $G_0 M G_0$ must be understood as matrix products. $G$ is diagonal in the basis of the Kohn-Sham bands, while $G_0$  is diagonal in the basis of the noninteracting bands. In general, however, both $G$ and $G_0$ are non-diagonal.  
 

 The self-energy is a  nonlinear functional of the Kohn-Sham propagator $G_{KS}$.  For our purposes it is sufficient to consider the first-order variation of the self-energy following from a small variation of the Green's function (the latter being induced by the magnetic perturbation):
 \be
 \delta M(\kv)= \sum_{\kv'}\frac{\delta M(\kv)}{\delta G_{KS,\kv'}}\delta G_{KS,\kv'}\,,
 \ee
 where the functional derivative is evaluated at the equilibrium state.  The dependence of $\delta M$ on $\delta G_{KS}$ arises from the density dependence of the effective potentials, which, in the linear response regime, are given by Eqs.~(\ref{DeltaVxc}) and (\ref{DeltaAxc}). 
 Fig.~\ref{Fig4}(b) shows the four diagrams that contribute to the linear variation of the Kohn-Sham self-energy.  Crucially, the functional derivative of $M$ with respect to $G_{KS}$ is given by 
 \be\label{WardIdentityText}
 \frac{\partial M(\kv)}{\partial G_{KS,\kv'}} = f_{n,n}+\fv_{n,\jv}\cdot\vv_{\kv'}+\vv_{\kv} \cdot\fv_{\jv,n}+\vv_\kv \cdot\stackrel{\leftrightarrow}{\fv}_{\jv,\jv}\cdot \vv_{\kv'}\,,
 \ee
 where $f_{n,n}$,  $\fv_{n,\jv}$ etc... without an argument are understood to be  the $\qv=0$ components of the corresponding kernels $f_{n,n}(\qv)$ etc..., $\vv_\kv$, $\vv_{\kv'}$ are the $\qv=0$ values of the  bare current vertex $\jv_\qv$ acting between states of Bloch wave vector $\kv$, $\kv'$ respectively. Notice that the kernels themselves do not depend on $\kv$
or $\kv'$, reflecting the locality of the Kohn-Sham potentials.

Inserting a bare current vertex $\hat\jv_{-\qv}$ on a Kohn-Sham propagator amounts to inserting it in turn on each one of the unperturbed propagators that appear in the expansion of the full propagator.  Whenever we do this, the unperturbed propagator is split into two unperturbed propagators that carry slightly different wave vectors $\kv$ and $\kv+\qv$.  
It is intuitively clear that the outcome of this procedure generates the  vertex correction diagrams depicted in Fig.~\ref{Fig3}. 

 \begin{figure}[ht]
\centering
\includegraphics[scale=0.5]{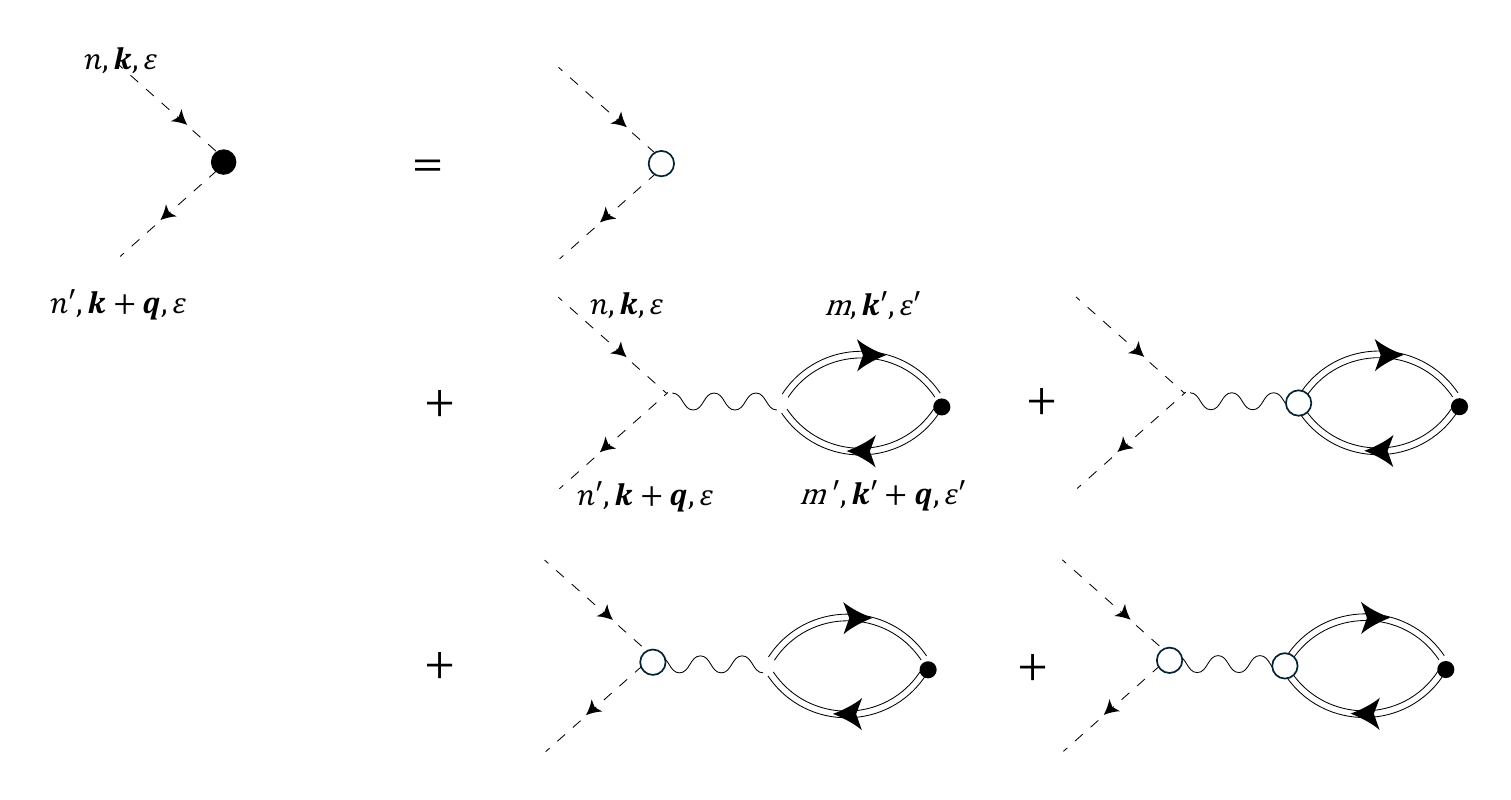}
\caption{\footnotesize{Diagrammatic representation of the integral equation for the dressed current vertex.  This follows directly from the diagrams of Fig.~\ref{Fig3}, after stripping away the energy density vertex on the left. Exchange-correlation kernels $f_{n,n}$, $\fv_{n,\jv}$, $\fv_{\jv,n}$ and $\stackrel{\leftrightarrow}{\fv}_{\jv,\jv}$ from left to right are represented by wavy lines. The internal labels $m,m',\kv',\epsilon'$ are summed over.}}
\label{Fig5}
\end{figure}

 We now prove that this sum of diagrams is equivalent to the single diagram of Fig. 2, in which the bare current vertex is replaced by the KS current vertex.  After noting that
  $\vv_\kv=-\partial [G_{0,\kv}]^{-1}/\partial \kv$ and $\vv^{KS}_\kv = -\partial [G_{KS,\kv}]^{-1}/\partial \kv$, use the Dyson equation
 \be
[G_{KS,\kv}]^{-1}=[G_{0,\kv}]^{-1}-M(\kv)\,.
 \ee
 to obtain
 \be\label{VKS-M}
 \vv^{KS}_\kv = \vv_\kv  + \frac{\partial M(\kv)}{\partial \kv}\,.
 \ee
 Notice that all quantities here are matrices with band and momentum indices, and all multiplications, inversions, etc.. must be performed following the standard rules of matrix algebra.  
 Then we notice that
 \be
 \frac{\partial M(\kv)}{\partial \kv} = -\sum_{\kv'}\frac{\partial M(\kv)}{\partial G_{KS,\kv'}} G_{KS,\kv'} \frac{\partial [G_{KS,\kv'}]^{-1}}{\partial \kv} G_{KS,\kv'} \,.
 \ee
 
 Inserting this into Eq.~(\ref{VKS-M}) with the help of Eq.~(\ref{WardIdentityText}) we get
 \be
 \vv^{KS}_\kv=\vv_\kv+\sum_{\kv'}\left\{f_{n,n}+\fv_{n,\jv}\cdot\vv_{\kv'}+\vv_{\kv} \cdot\fv_{\jv,n}+\vv_\kv \cdot\stackrel{\leftrightarrow}{\fv}_{\jv,\jv}\cdot \vv_{\kv'}\right\}G_{KS,\kv'} \vv^{KS}_{\kv'} G_{KS,\kv'}\,.
 \ee
  This coincides formally with the equation for the dressed current vertex, which is displayed in Fig.~\ref{Fig5}, proving that the dressed current vertex at $\qv=0$ is indeed the KS velocity vertex. We emphasize that our result is valid only for the $\qv\to 0$ limit of the dressed current vertex, but this is all we need, since the linear in $\qv$ corrections to the current vertex do not contribute to the magnetization, as discussed at the end of Section~\ref{NonInteractingSystem}.

\section{Extension to finite temperature}
Thus far, our analysis has been restricted to the zero-temperature limit. However, the derivation can be repeated at finite temperature, where the grand-canonical energy density is replaced by the grand-canonical thermodynamic potential density, which includes an entropy density contribution.  
To this end, we  closely follow the Mermin formulation of finite-temperature DFT\cite{Mermin1965,Kohn1983} and write the grand-canonical potential (at equilibrium)  as
\ber\label{EquilibriumPhi}
\Phi&=& \sum_{n,\kv} \left\{f_{n,\kv}\left[\epsilon^{(0)}_{n,\kv}-\mu\right] +\beta^{-1}\left[f_{n,\kv}\ln f_{n,\kv}+(1-f_{n,\kv})\ln(1-f_{n,\kv})\right]\right\}\nn\\
&+& \Phi_{Hxc}[n,\jv_p]\,.
\eer
{Here $\epsilon^{(0)}_{n,\kv}$ are  the  expectation values of the noninteracting Hamiltonian $\hat T_\Av+\hat V$ in the exact eigenstates of the Kohn-Sham equation  ($\hat T_\Av \equiv \frac{1}{2}\left[-i\hat\nablav_\rv+e\Av(\hat\rv)\right]^2$ is the gauge-invariant kinetic energy operator, $\hat V \equiv V(\hat\rv)$ is the external potential operator).  The Fermi-Dirac occupation factors $f_{n,\kv}$ are computed from the  eigenvalues $\epsilon_{n,\kv}$  of the Kohn-Sham Hamiltonian 
\be\label{KohnShamHamiltonian}
\hat H_{KS}\equiv \hat T_\Av+\hat V+ \frac{1}{2}\left\{-i \hat \nablav_\rv,\cdot \Av_{xc}(\hat\rv)\right\}+ V_{Hxc}(\hat \rv)
\ee
 with xc potentials given as functional derivatives of the temperature-dependent  Hartree+xc grand-potential functional $\Phi_{Hxc}[n,\jv_p]$ with respect to the densities $n$ and $\jv_p$,   evaluated at the equilibrium values of these densities. 
 The first line of Eq.~(\ref{EquilibriumPhi}) is the noninteracting grand-canonical potential functional generated by the Kohn-Sham orbitals. The second line is the interaction contribution\cite{Mermin1965,Kohn1983}. Notice that Eq.~(\ref{EquilibriumPhi}) is the finite-temperature extension of Eq.~(\ref{EnergyFunctional}), as can be verified by taking the zero-temperature ($\beta\to \infty$) limit. We emphasize that this expression is equivalent to Eq.(18) of Ref.~\cite{Shi2007}.}

For small deviations from equilibrium, we linearize $\Phi_{Hxc}[n,\jv_p]$ about the equilibrium densities.
Recall that
\be
\delta  \Phi_{Hxc}[n,\jv_p] = \int \left\{V_{Hxc}(\rv)\delta n(\rv)+\Av_{xc}(\rv)\cdot \delta \jv_p(\rv)\right\}d\rv\,.
\ee
This allows us to ``absorb" the variation of $\Phi_{Hxc}$  (near equilibrium) in a one-body term controlled by the KS Hamiltonian~(\ref{KohnShamHamiltonian}) as follows
\be
\delta \Phi =  \sum_{n,\kv}\bar \delta  \left\{f_{n,\kv}\langle\psi_{n,\kv}|\hat H_{KS}-\mu|\psi_{n,\kv}\rangle  +\beta^{-1}\left[f_{n,\kv}\ln f_{n,\kv}+(1-f_{n,\kv})\ln(1-f_{n,\kv})\right]\right\}\,.
\ee
As in Eq.~(\ref{deltaE2}),  there is one important caveat: namely,  the Kohn-Sham Hamiltonian operator $\hat H_{KS}$ is  pinned at its equilibrium value (this is the meaning of the notation $\bar \delta$). The  KS eigenfunctions $|\psi_{n,\kv}\rangle$, on the other hand, as well as their eigenvalues $\epsilon_{n,\kv}$, hidden in the occupation factors $f_{n,\kv}$,  are functionals of the varying densities: their variation is computed from the solution of a Kohn-Sham equation which  include the effect of varying Hxc potentials. With this understanding, a dramatic simplification occurs and the grand-canonical potential can be expressed as
%
%
\be
 \delta \Phi= \int d\rv ~\bar \delta \left\{ \left( -\beta^{-1}\sum_{n,\kv}\ln\left[1+e^{-\beta(\epsilon_{n,\kv}-\mu)}\right]  \psi^*_{n,\kv}(\rv)\psi_{n,\kv}(\rv)\right)\right\}\,.
\ee
This is the extension of Eq.~(\ref{deltaE2}) to finite temperature. By the principle of nearsightedness~\cite{Prodan2005} the integrand of the above expression is interpreted as the grand-thermodynamic potential density. 
Its  Fourier component at wave vector $\qv$ is
\be\label{PhiQ}
 \delta \Phi_\qv = -\beta^{-1}\sum_{n,\kv}\bar \delta \langle\psi_{n,\kv}|\ln\left[1+e^{-\beta(\hat H_{KS}-\mu)}\right]  \hat n_\qv|\psi_{n,\kv}\rangle\,.
\ee

The sum over states in this equation has the structure of a trace in the Hilbert space of Bloch wave functions. 
When a periodic magnetic field of amplitude $\Bv_\qv$ is applied the value of the trace can change in two ways. One is through the explicit addition of the external magnetic field to $\hat H_{KS}$. The other is through a spectral flow that changes the distribution of the Bloch states in $\kv$-space. However, the first mechanism is not viable because the expectation value of the periodic perturbation vanishes in the thermal ensemble of unperturbed Bloch wave functions.
The orbital magnetization $\Mv= -\lim_{\qv\to 0}\delta\Phi_\qv/\delta \Bv_\qv$ is therefore entirely controlled by the spectral flow mechanism. 

The essential point here is that the perturbed Bloch wave functions are eigenfunctions of a perturbed Kohn-Sham Hamiltonian in which the effective potentials are allowed to change along with the external magnetic field. 
We have shown in Section~\ref{WardIdentity} that in this process the bare current vertex $\vv_\kv$ (the open dot in Fig.~\ref{Fig5}) is ``dressed" by the variation of the xc potential and morphs to the Kohn-Sham velocity vertex $\vv_\kv^{KS}$ (the black dot in Fig.~\ref{Fig5}).  From this it follows that the spectral flow of the Kohn-Sham wave functions in Eq.~(\ref{PhiQ}) is entirely controlled by the Kohn-Sham velocity, as it would be the case  if the Kohn-Sham Hamiltonian were interpreted as a genuine non-interacting Hamiltonian. Therefore the calculation of the spectral flow proceeds exactly as in the noninteracting case, and the result is given by the finite-temperature magnetization formula~(\ref{ModernTheory}) with all quantities computed from the Kohn-Sham eigenfunctions and eigenvalues.

\section{Concluding remarks}

This work reasserts, through a much more careful analysis, the claim that was made in Ref.~\cite{Shi2007}: the orbital magnetization of an interacting periodic solid can be calculated via the ``modern theory" formula, Eq.~(\ref{ModernTheory}), with Bloch functions and band energies obtained from the self-consistent solution of the Kohn-Sham equation of CDFT.

There are two new ingredients in this work. 
First, we make use of CDFT to calculate the response of the energy density -- not the total energy -- to a periodic magnetic field in the long-wavelength limit. Notice that this aligns with the first part of Ref.~\cite{Shi2007}, where the non-interacting formula was indeed calculated from the response of the energy density. Second, we recognize the key role played by a Ward identity, Eq.~(\ref{WardIdentityText}), which connects the long-wavelength limit of the exchange-correlation kernels (derivatives of the potentials with respect to densities) to derivatives of the Kohn-Sham self-energy. This identity is a new result of this work.  Physically, it asserts that the magnetic perturbation couples to the Kohn-Sham system via the dressed velocity operator $\vv_{\kv}^{KS}=\partial\hat H_{KS,\kv}/\partial\kv$, rather than the bare one $\vv_{\kv}=\partial\hat H_{\kv}/\partial\Av$.

Our findings are perfectly aligned with recent results proving the applicability of the orbital magnetization formula in Hartree-Fock theory~\cite{Vafek2025,Chunli2025,Xiao2025}. We believe that whenever a many-body system is approximately described by a self-consistent single-particle Hamiltonian the orbital magnetization formula provides a consistent way to calculate the orbital magnetization {\it within that approximation}.


What sets CDFT apart from mean field theories is its claim to ``formal exactness".  If the exact xc energy functional were known, then the calculated magnetization would not be an approximation, but the true magnetization of the interacting system.
CDFT stands alone in this claim.   Ordinary DFT~\cite{Grayce94} -- in spite of being formally exact for the ground state energy and the density -- cannot make it, because its Kohn-Sham orbitals do not yield the exact current density. This does not mean that an exact calculation of the orbital magnetization within ordinary DFT is  impossible. But it means that such a calculation cannot be done using only the modern theory formula: it would be necessary to add a correction arising from the derivative of the (exact) exchange-correlation energy functional with respect to the magnetic field.

{As a final point, we reiterate that our analysis is firmly based on the {\it energy-density} approach to the calculation of the orbital magnetization. An alternative approach, in which one computes the total energy of a large but finite crystallite subjected to a uniform magnetic field is  possible~\cite{Thonhauser2005,Ceresoli2006}, but requires explicit consideration of the edge structure of the crystallite. Since the orbital magnetization is  widely believed~\cite{BiancoResta13} to be a bulk property, we expect that the total energy calculation, if properly performed, will give the same result as the energy-density calculation, even in the presence of interactions.}

\begin{acknowledgements}
    GV was supported by the Ministry of Education, Singapore, under its Research Centre of Excellence award to the Institute for Functional Intelligent Materials (I-FIM, project No. EDUNC-33-18-279-V12). JS was supported by the National Science Foundation of China, Grant No. 12574169. QN was supported by work was supported by the National Natural Science Foundation of China (No. 12234017) and National Key R\&D Program of China (No. 2023YFA1407001). Special thanks are due to Jacques Desmarais and Stefano Pittalis for starting the discussion that led us to revisit the derivation of the magnetization formula in CDFT.
\end{acknowledgements}

\bibliographystyle{apsrev4-2}
\bibliography{CDFT-Orbital.bib}

\end{document}